\documentclass[prd,floatfix,preprintnumbers]{revtex4}
\usepackage{amssymb}
\usepackage{amsmath}
\usepackage{graphicx}
\usepackage{dcolumn}

\newcommand {\ga} {\ {\raise-.5ex\hbox{$\buildrel>\over\sim$}}\ }
\newcommand {\la} {\ {\raise-.5ex\hbox{$\buildrel<\over\sim$}}\ }

\begin{document}
\title{Gravitational Effects of a Small Primordial Black Hole Passing Through the Human Body}
\author{Robert J. Scherrer}

\affiliation{Department of Physics and Astronomy, Vanderbilt
  University, Nashville, TN, 37235, USA}
	
\begin{abstract}The gravitational effects of a primordial black hole (PBH) passing through the human body are examined, with
the goal of determining the minimum mass necessary to produce significant injury or death.  Two effects are examined:
the damage caused by a shock wave propagating outward from the black hole trajectory, and the dissociation of brain
cells from tidal forces produced by the black hole on its passage through the human body.  It is
found that the former is the dominant effect, with a cutoff mass for serious injury or death of approximately
$M_{PBH} > 1.4 \times 10^{17} {\rm g}$.  The number density of primordial black holes
with a mass above this cutoff is far too small to produce any observable effects on the human population.
\end{abstract}

\maketitle

\section{Introduction}
The detection of gravitational radiation from merging black holes \cite{LIGO} as well as the direct imaging of
M87* \cite{M87*} and Sgr A* \cite{SgrA*} has led to a significant resurgence of interest in black hole physics.  Among
the ideas that have been reexplored is the possibility that primordial black holes (PBH) could account for some or all of the dark matter
in the Universe \cite{Carr(1974),Chapline(1975),Carr(1975),Khlopov(2010),Belotsky(2014),Carr(2020)}.  Many constraints can be placed on the mass of black holes
that can serve as dark matter; the main open mass window lies in the asteroid mass range \cite{Carr(2021),Gorton(2024)}
\begin{equation}
\label{limits}
10^{17} {\rm g} < M_{PBH} < 10^{22} {\rm g}.
\end{equation}
Here Hawking radiation provides the lower bound: any PBHs with a mass below $10^{15}$ g would have evaporated by
the present day, while limits on the PBH decay products push this lower bound up to about $10^{17}$ g.
The upper
bound in Eq. (\ref{limits}) is due to limits on black hole microlensing events, but other limits rule out black hole dark matter at higher
masses.  Any black holes lying in the mass range given by Eq. (\ref{limits}) must be primordial, as they cannot
be produced through normal stellar evolution.

If primordial black holes exist and are sufficiently abundant, they could produce a variety of interesting effects.
Several authors have explored the consequences of a PBH impact on the Earth
\cite{Khriplovich(2008),Luo(2012),Rahvar(2021)}, and
in Ref. \cite{Luo(2012)} it was argued that a PBH as small as $10^{15}$ g would produce
an easily-detectable seismic disturbance.  Dai and Stojkovic \cite{Dai} have suggested that PBHs could accumulate
inside of planets and asteroids and hollow out their cores.  They also pointed
out that the passage of a PBH through material on the earth would
leave a detectable microscopic tunnel and suggested further that such a passage through
the human body would not be fatal.  It is this latter question that we address here.

Clearly, a sufficiently large
PBH would have a significant impact on the human body; the goal of this present work is to determine
the minimum black hole size needed to produce significant injury, and to see whether the observed lack of such injuries
can place any bounds on PBH properties.  Our discussion is partially motivated by earlier work on macroscopic dark matter (MACROs) \cite{Sidhu(2020)}.
In that paper, it was shown that the impact of MACRO dark matter on the human body would be sufficiently destructive that the non-observance of such impacts allows useful limits to be placed
on the MACRO mass and cross section.   While
it is well known that a microscopic black hole would have a deleterious effect on the human body,
there are no precise calculations.  Most previous discussions of this topic are in the popular literature
\cite{Loeb(2021)}; in fact, the first reference to the effect of a PBH on the human body
appears to be in a work of fiction \cite{Niven(1974)}.  Lemos \cite{Lemos(1996)} noted that very small black holes (less massive
than those considered here) would have essentially no gravitational effect on the human body, but that the
Hawking radiation could be quite destructive.  Here we consider only gravitational forces,
leaving the discussion of Hawking radiation for future work.

In the next section, the biological effects of a PBH passing through the human body are discussed, including both the shock wave produced by such a passage and
the tidal forces that would tear apart human cells.  The minimum mass needed to produce serious injury or death from each of these effects is derived.
In Sec. III we examine whether such effects can place useful limits on PBH properties.  While the minimum mass needed to produce
serious injury or death lies within the interesting asteroid mass range, the allowed PBH abundance in this mass range is too small to produce any observable
effects on the human population. 

\section{Primordial black holes and the human body}

Here we will consider two different types of gravitational effects that would be produced by
a PBH collision with the human body.  First, the PBH would generate
a supersonic shock wave along its path, destroying tissue along the way.
Second, tidal gravitational forces would tend to tear apart cells in the body.
In our mass range of interest (Eq. \ref{limits}), the Schwarzschild radius is $R_S \la 10^{-6}$ cm,
so we are justified in treating the PBH as a point mass, even at the cellular level.

Consider first the shock wave generated by a PBH passing through the human body.
Khripolovich et al. \cite{Khriplovich(2008)} estimated the total energy that would be deposited by
a PBH in the form of a shock wave as it passed through the Earth.  Their estimate for the total
energy deposition is \cite{Khriplovich(2008)}
\begin{equation}
\label{Edeposit}
E = 4 \pi (GM_{PBH}/v_{PBH})^2 \rho L \ln(c_s/\sqrt{4 \pi G \rho} a).
\end{equation}
Here $M_{PBH}$ and $v_{PBH}$ are the PBH mass and velocity, $\rho$ is the density of the target,
$L$ is the path length of the PBH through the target, $c_s$ is the sound speed in the target,
and $a$ is the typical intermolecular distance in the target.  We can take the human body to be primarily water, with $\rho = 1$ g/cm$^3$, $c_s = 1.5 \times 10^5$ cm/s, and $a = 3 \times 10^{-8}$ cm.
Then the logarithmic factor on the right-hand side of Eq. (\ref{Edeposit}) is 36.

How much energy would need to be deposited in the human body to produce significant injury or death?  The shock wave produced in this process would be similar to that produced by a bullet passing through the human body, which
is a well-studied process.  In the latter case, the
tissue damage scales roughly as the deposited energy \cite{Courtney(2012),Baum(2022)}.  Following
Ref. \cite{Sidhu(2020)}, we will take the threshold for serious energy to be approximately
100 J, which is the typical muzzle energy of a .22 caliber rifle \cite{Rhee(2016)}.  Using this value for $E$ in Eq. (\ref{Edeposit}), along with a typical path length
through the human body of $L =$ 10 cm and a PBH velocity on the order of the
dark matter velocity dispersion ($v_{PBH} \sim 200$ km/s), we find
that the lower bound on $M_{PBH}$ to produce significant injury in this fashion is
\begin{equation}
\label{shocklimit}
M_{PBH} > 1.4 \times 10^{17} {\rm g}.
\end{equation}

Now consider a second mechanism for tissue destruction. As the PBH passes
through the human body, it will exert strong tidal forces in its vicinity. 
These will produce a tensile force on nearby human cells, and a sufficiently strong
force would tear the cells apart.  The cells most sensitive to this dissociation are likely to be those
in the human brain.  A tensile force of $10-100$ nN for a few microseconds would be sufficient to
pull apart human brain cells \cite{Miller(2002),Hutson(2024)}.

Recall that the tidal force 
on a cell with mass $m_{cell}$ and typical size $r_{cell}$ at a distance $d$ from the PBH is given by
\begin{equation}
F = \frac{2GM_{PBH} m_{cell} r_{cell}}{d^3}.
\end{equation}
Note that the brain can be surprising resilient to damage over very small regions.  Hence, we will err on the side of conservatism
and require that the entire brain be subject to a tidal force of $10-100$ nN.  This will insure sufficient
destruction of brain tissue (basically, the entire brain) to produce a detectable effect. 
Furthermore,
a PBH velocity of ~200 km/s is equivalent to 20 cm/$\mu$s, so that
it will take $\sim \mu$s for the PBH to traverse the entire brain, ensuring that these
tidal forces will be applied to the entire brain for at least a few microseconds.
We will take $r_{cell} \sim 10 ~\mu$m, and $m_{cell} \sim \rho r_{cell}^3$ with $\rho$ again taken
to be the density of water.  Then with $d \sim 10$ cm,
we find that $F \sim 10-100$ nN corresponds to a lower bound on the PBH mass of
\begin{equation}
M_{PBH} > 7 \times 10^{18} {\rm g} - 7 \times 10^{19} {\rm g}.
\end{equation}
Even given the crudeness of our approximations, it is clear that
the effect of the gravitationally-induced shock wave on the human body is larger than the effect of the tidal
forces. Thus, Eq. (\ref{shocklimit}) gives the smallest PBH mass that will produce significant damage to the human body
from gravitational forces.

\section{Conclusions}

Intriguingly, the smallest PBH mass that can lead to significant human injury (Eq. \ref{shocklimit}) lies near the lower bound on PBH dark matter in Eq.
(\ref{limits}), raising the possibility
that that the absence of such events could provide an additional constraint on the PBH abundance, as in the case of MACHO dark matter
\cite{Sidhu(2020)}.  Unfortunately, it is easy to see that the effect of PBH collisions
with the human population will always be negligible.  Following Ref. \cite{Sidhu(2020)},
we estimate the rate of PBH-human interactions as
\begin{equation}
\Gamma = \frac{\rho_{PBH} N_{human} A_{human} v_{PBH}}{M_{PBH}},
\end{equation}
where $N_{human}$ is the total human population ($N_{human} = 8 \times 10^9$) and $A_{human}$ is the
typical cross-sectional area of a human ($A_{human} \sim 1 ~{\rm m}^2$).
Taking $\rho_{PBH}$ to account for all of the dark matter ($\rho_{PBH} = 2 \times 10^{-30}$ g/cm$^{3}$), and using the smallest mass in Eq.
(\ref{shocklimit}) that allows for serious injury, we obtain $\Gamma \sim 10^{-18}$ events per year, which is utterly negligible.

This result is not suprising, as there are two major differences between PBHs and the model considered in Ref. \cite{Sidhu(2020)}.  The first
is that the MACHO model allows for much smaller dark matter masses and correspondingly larger number densities, resulting
in a higher interaction rate with the human population.  The second is that MACHOs
are strongly interacting, while PBHs interact gravitationally, and gravity is, of course,
notoriously weak.  Perhaps the more interesting result is just how weak the limit in Eq.
(\ref{shocklimit}) turns out to be.  A black hole with $M_{PBH} \sim 10^{16} $ g corresponds to the mass of a small
asteroid, but such a black hole could pass right through the human body and cause only negligible
gravitational effects.

\begin{acknowledgments}

I thank A. Loeb, M.S. Hutson, and J.R. Scherrer for helpful discussions.
\end{acknowledgments}

\end{document}